\documentclass[accepted]{article}

\usepackage{microtype}
\usepackage{graphicx}
\usepackage{subfigure}
\usepackage{booktabs} 
\usepackage{caption}

\usepackage{hyperref}

\usepackage{icml2022}

\usepackage{amsmath}
\usepackage{amssymb}
\usepackage{mathtools}
\usepackage{amsthm}

\usepackage[capitalize,noabbrev]{cleveref}

\theoremstyle{plain}

\theoremstyle{definition}

\theoremstyle{remark}

\usepackage[textsize=tiny]{todonotes}
\usepackage[normalem]{ulem}

\usepackage{caption}
\usepackage{dblfloatfix}

\newcommand{\hbadd}[1]{\textcolor{blue}{#1}}

\icmltitlerunning{Identifying Orientation-specific Lipid-protein Fingerprints using Deep Learning}

\begin{document}

\twocolumn[%
\icmltitle{Identifying Orientation-specific Lipid-protein Fingerprints using Deep Learning}


\icmlsetsymbol{equal}{*}

\begin{icmlauthorlist}
\icmlauthor{~~Fikret Aydin}{pls}
\icmlauthor{~~Konstantia Georgouli}{pls}
\icmlauthor{~~Gautham Dharuman}{pls}
\icmlauthor{~~James N Glosli}{pls}
\icmlauthor{~~Felice C Lightstone}{pls}
\icmlauthor{~~Helgi I Ing\'olfsson}{pls}
\icmlauthor{~~Peer-Timo Bremer}{comp}
\icmlauthor{~~Harsh Bhatia}{comp}
\end{icmlauthorlist}

\icmlaffiliation{pls}{Physical and Life Sciences, Lawrence Livermore National Laboratory, Livermore, California, 94550}
\icmlaffiliation{comp}{Center for Applied Scientific Computing, Lawrence Livermore National Laboratory,Livermore, California, 94550}

\icmlcorrespondingauthor{Fikret Aydin}{aydin1@llnl.gov}

\icmlkeywords{Machine Learning, ICML}

\vskip 0.3in
]



\printAffiliationsAndNotice{\icmlEqualContribution} 

\begin{abstract}
Improved understanding of the relation between the behavior of RAS and RAF proteins and the local lipid environment in the cell membrane is critical for getting insights into the mechanisms underlying cancer formation. In this work, we employ deep learning (DL) to learn this relationship by predicting protein orientational states of RAS and RAS-RAF protein complexes with respect to the lipid membrane based on the lipid densities around the protein domains from coarse-grained (CG) molecular dynamics (MD) simulations. Our DL model can predict six protein states with an overall accuracy of over 80\%. The findings of this work offer new insights into how the proteins modulate the lipid environment, which in turn may assist designing novel therapies to regulate such interactions in the mechanisms associated with cancer development.

\end{abstract}

\vspace{-1em}
\section{Introduction}

RAS proteins play a key role in cellular mechanisms that usually involve regulation of complex protein-protein and protein-lipid interactions. Imbalance in these regulatory mechanisms can result in malfunction of cells and occurrence of serious diseases. Mutations in RAS proteins are responsible for the development of about a third of all human cancers~\cite{simanshu_ras_2017, Prior2969}. Complex protein-lipid interactions at the cell membrane can be studied using molecular dynamics (MD) simulations to understand the role of RAS proteins in cancer formation.

The behavior of RAS proteins is known to be mediated by membrane lipid composition. Here, we are interested in understanding this phenomenon in the form of \textit{``lipid fingerprints''}, which are the spatial distributions of lipids in the membrane in response to  proteins~\cite{Corradi18}. We are specifically interested in understanding lipid fingerprints that are associated with unique protein orientational RAS states with respect to the membrane because lipid fingerprints represent complex lipid-protein interactions, and such specific types of fingerprints may be used to modulate RAS signaling effectiveness, potentially informing of ways to intercept this modulation.

Recent work~\cite{ingolfsson2022machine} has shown that RAS creates specific lipid fingerprints and that deep learning (DL) can be used to predict RAS protein orientational states with respect to the membrane based on the fingerprints. However, their study focused on single-protein systems that include only RAS. Here we focus on more-complex systems and investigate if distinct lipid fingerprints exist for RAS-RAF complexes as the behavior of these complexes is still not well understood.

State prediction becomes more challenging with the inclusion of RAS-RAF protein complexes due to more complex protein-lipid interactions from new electrostatic and hydrophobic interactions of the RAF protein domains. In addition, there is a competition between RAS and the RAF domains to interact with available lipid molecules around proteins, which can add significant noise to the data as lipid concentrations around one protein may be perturbed by the molecular interactions formed by the other protein. The prediction is  challenging also because even small differences in the lipid densities can affect the protein state.

Here, we present a new DL-based framework to predict protein orientational states of RAS proteins as well as RAS-RAF protein complexes from spatial distribution of lipids. The data is gathered from on coarse-grained (CG) molecular dynamics (MD) simulations that are part of a larger multiscale simulation~\cite{mummi:SC:2021}. By considering protein states as labels and lipid densities around the protein as features, identification of lipid fingerprints can be cast into a standard classification problem. Specifically, we use a deep neural network based on the ideas of residual networks (ResNets)~\cite{resnet_2015}. ResNets have been shown to be powerful tools for image classification as they can capture various features in images~\cite{resnet_2015, facerecognition}. Such models are particularly attractive because they facilitate deeper models without overfitting and, hence, allow learning complex features with better accuracy than ordinary convolutional neural networks.

Using a total of 2,400,000 MD snapshots --- each containing either a RAS protein or a RAS-RAF complex along with 14 lipid species --- we train and validate our model to predict six different protein orientational states. We demonstrate an overall accuracy of over 80\% and present an initial view on the representative fingerprints associated with these states. Our results provide important insights into the effect of local lipid environment on RAS-RAF behavior, which can shed light on the mechanisms underlying cancer formation.

\section{Methods}

\subsection{Description of Data} 
The data used here was generated from a recent multiscale simulation~\cite{mummi:SC:2021}. From the entire simulation set, we used 6420 MD simulations~\cite{Zhang_ddcMD_2020} containing either one RAS protein or one RAS-RAF complex. These simulations amounted to 2,400,000 MD snapshots saved at a temporal resolution of 10 ns. To provide a consistent frame of reference for lipid fingerprints, the snapshot was transformed such that it is centered around the farnesyl and the G-domain aligned with the positive $x$-axis (both farnesyl and G-domain are components of the RAS protein). 

Since the fingerprints concern with the spatial distribution of lipids, and not with the individual lipids themselves, we cast the particle representation of lipids into lipid densities using kernel density estimation. Density of lipids within the 30$\times$30 nm$^2$ simulation box was estimated on a 37$\times$37 grid and saved as an image. In molecular dynamics simulations, a system is commonly referred to as a “simulation box”, which is prepared by defining a bounded volume and inserting a certain number of different types of molecules such as lipids and proteins into this volume. A 13$\times$13 subgrid (equivalent to 10$\times$10 nm$^2$) from the center was extracted because lipid-protein fingerprints exist only in a small neighborhood around the protein. Densities of the 14 lipids were treated independently, resulting in a 13$\times$13$\times$14 multichannel image. 
The multichannel images were standardized per channel so that they have zero mean and unit standard deviation across the dataset to transform ranges of different lipid concentrations into comparable scales.
As documented previously~\cite{ingolfsson2022machine}, the tilt and rotation angles of the RAS G-domain with respect to the lipid membrane and distance between CRD region of protein and membrane (used only for RAS-RAF) were used to describe the protein orientational states of RAS and RAS-RAF. A total of six states were identified in the dataset (along with their population, which are stated in paranthesis): $\alpha$ (10.9\%), $\beta$ (24.6\%), $\beta'$ (7.6\%), \texttt{ma} (26.4\%), \texttt{mb} (27.6\%), and \texttt{za} (2.9\%). Extensive CG simulations have been used to identify these canonical protein orientational states~\cite{ingolfsson2022machine}.

\begin{figure*}[!t]
\centering
\includegraphics[width=0.75\linewidth]{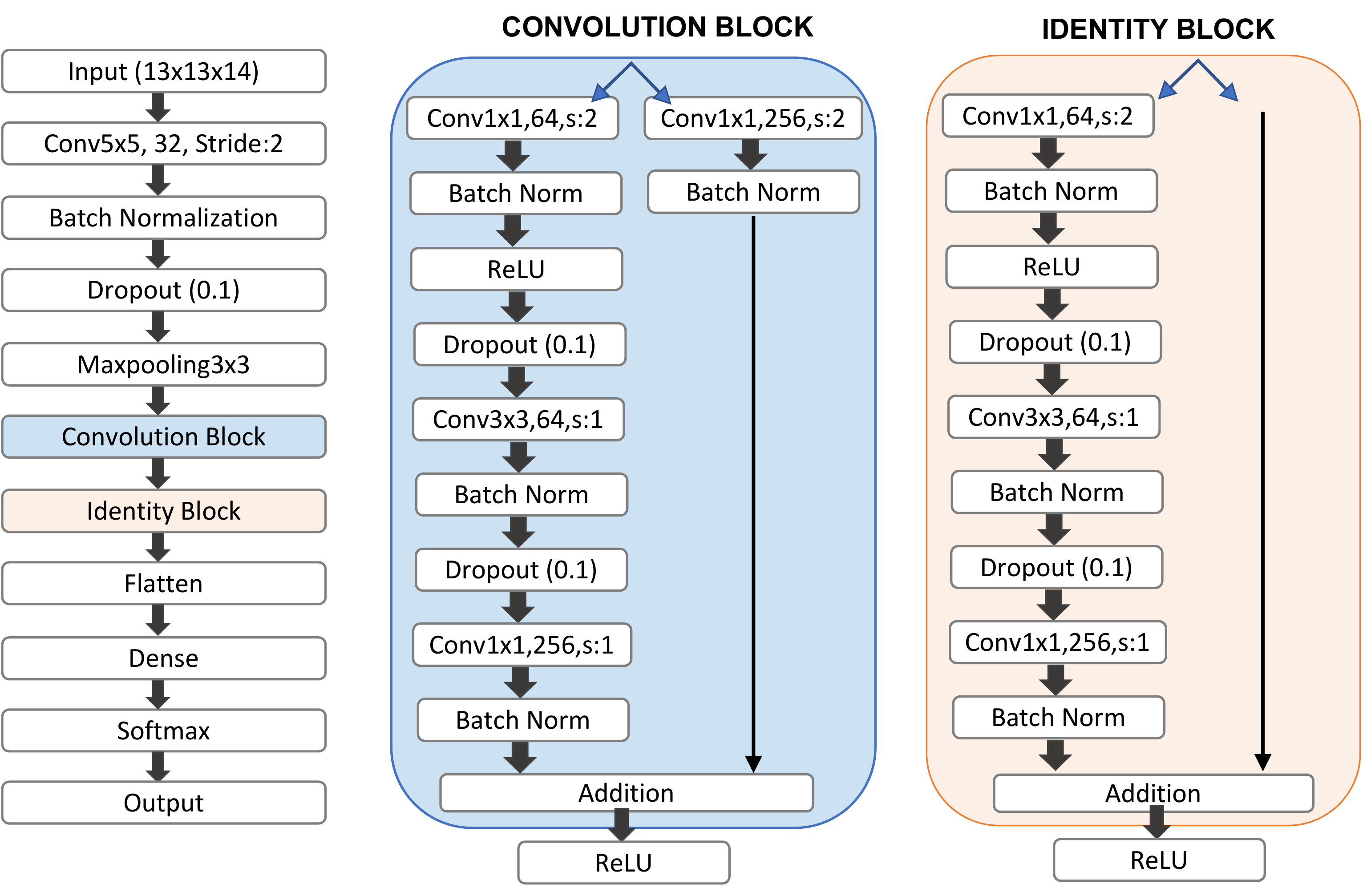}
\vspace{-0.75em}
\caption{Our DL architecture is inspired by the ResNet approach that uses ``skip'' connections leading to the ability to capture more complexity without overfitting. Our model uses one convolution and one identity block as incorporated within a feed-forward network.}
\vspace{-1.25em}
\label{fig:model}
\end{figure*}

\subsection{Description of the Deep Learning Model}

Given the input multichannel images $X$ (lipid densities on a 13$\times$13$\times$14 grid) and the associated labels (six possible protein states), our goal is to learn a mapping $X\to Y$ that allows us to predict the protein state, given the lipid densities. To solve this classification problem, we develop a deep neural network inspired by the ResNet approach.

Illustrated in \autoref{fig:model}, our model comprises a convolutional block and an identity block, which are incorporated into a feed-forward network. We make use of batch normalization layers and \textit{relu} activations with the convolutional layers --- benefits of both are well established in the literature. We also note that insertion of dropout layers helped with preventing overfitting. Using this model, we reduced the data to a 6-dimensions (corresponding to six different protein states) predicting the unscaled log probabilities (logits), which are then transformed with \textit{softmax} to create class probabilities.

The given data was split 90\%--10\% into train-validation sets. The model was optimized using the categorical cross-entropy loss with the Adam optimizer~\cite{kingma2014adam} and a batch size of 4096 and learning rate 0.0005. The model was trained until the loss appeared converged (about 500 epochs). The model was built using the TensorFlow~\cite{tf} and Keras~\cite{keras} libraries. A data-parallel approach was employed using the Horovod framework~\cite{horovod} to distribute training over 10 computational nodes, where a copy of the same model is trained by each node. Each computational node contains four NVIDIA Volta 100 GPUs.

\section{Results}

We used our framework to train and validate our model as described above using the given 2,400,000 snapshots.

\begin{figure}[!t]
\centering
\vspace{-1.5em}
\includegraphics[height=2.65in]{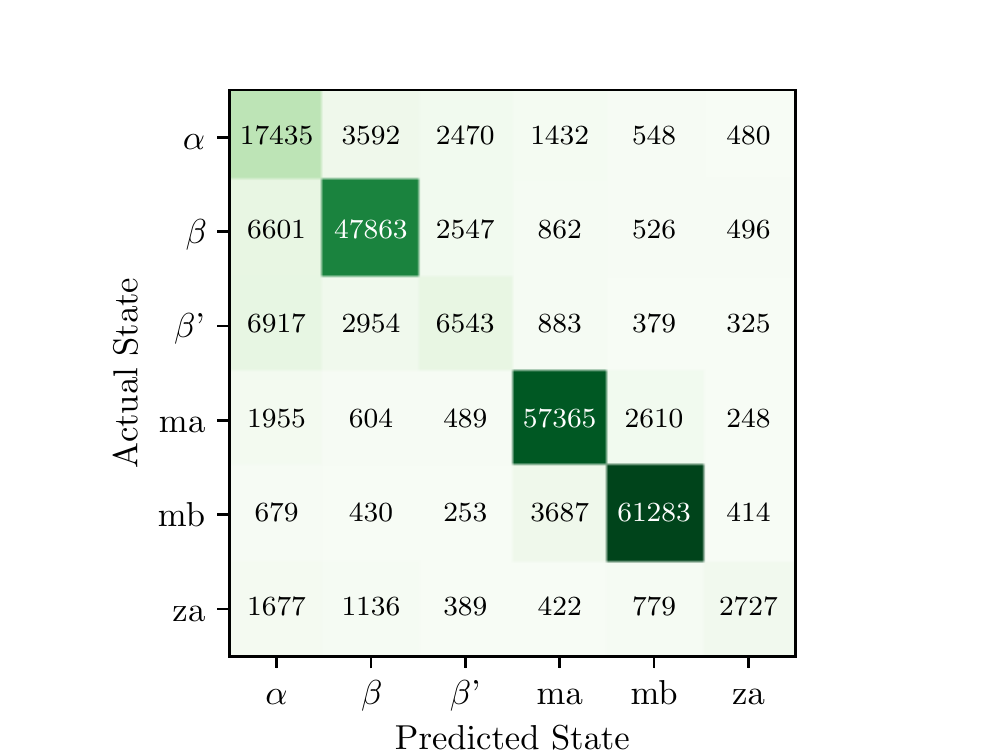}
 \vspace{-2em}
\caption{The confusion matrix for the presented model highlights good classification capability, and that certain states are more difficult to predict than others. The tilt and rotation angles of the RAS G-domain with respect to the lipid membrane and distance between CRD region of protein and membrane (used only for RAS-RAF) were used to describe the protein orientational states of RAS and RAS-RAF. Tilt angles define the tilt of helix 5 region of RAS protein relative to the lipid membrane normal, and rotation angles define the direction of that tilt.
\vspace{-1.5em}}
\label{fig:confmat}
\end{figure}

Here, we show the classification performance of our model using the confusion matrix in \autoref{fig:confmat}. A confusion matrix summarizes the results of a classification model by summarizing the total number of correct and incorrect classifications --- the rows of a confusion matrix indicate the true class labels whereas the columns indicate the predicted class labels. Using the confusion matrix, it is straightforward to count the true positives (TP), true negatives (TN), false positives (FP), and false negatives (FN). We further distill these results into the commonly-used metrics of accuracy, precision, and recall for classification (see \autoref{tab:apr}). Accuracy is calculated as the ratio of all correct predictions to all predictions, i.e., (TP+TN)/(TP+TN+FP+FN); precision measures how many data points of a predicted class actually belong to the said class, i.e., TP / (TP+FP), and recall measures how many data points within a given class were predicted correctly, i.e., TP / (TP+FN).

\begin{table}[!t]
\centering
\vspace{-0.5em}
\caption{Summary of the accuracy, precision, and recall of our classification model.\label{tab:apr}}
\vspace{0.5em}
\begin{small}
\begin{tabular}{lrrr}
\toprule
State      & Accuracy & Precision & Recall \\
\toprule
$\alpha$   & 67.17 & 48.75 & 64.88 \\
$\beta$    & 81.27 & 82.24 & 81.90 \\
$\beta'$   & 36.35 & 48.00 & 33.93 \\
\texttt{ma} & 90.67 & 86.14 & 87.86 \\
\texttt{mb} & 91.82 & 88.71 & 91.87 \\
\texttt{za} & 38.25 & 53.20 & 31.45 \\
\midrule
Overall    & 80.51 & 80.05 & 80.07 \\
\bottomrule
\end{tabular}
\end{small}
\vspace{-1.25em}
\end{table}

The results indicate that the model performance depends on the type of protein orientational state. Whereas some states were predicted with an accuracy of over 90\%, others were predicted with only about 35-40\%. In particular, this gap in performance is likely due to class imbalance issues, since the protein states with lower accuracy correspond to those with small amounts of training data. In addition, some protein orientational states, such as \texttt{za}, have a smaller membrane footprint as the important membrane binding regions of the proteins are located away from the lipid membrane, which can lead to the low accuracy due to lack of extensive protein-lipid interactions. Although thus far, we chose to explore the data without augmentation, these initial results demonstrate the need for data augmentation --- an additional step we wish to incorporate in the next version of the model.

\begin{figure}[!t]
\centering
\includegraphics[width=1.0\linewidth,keepaspectratio]{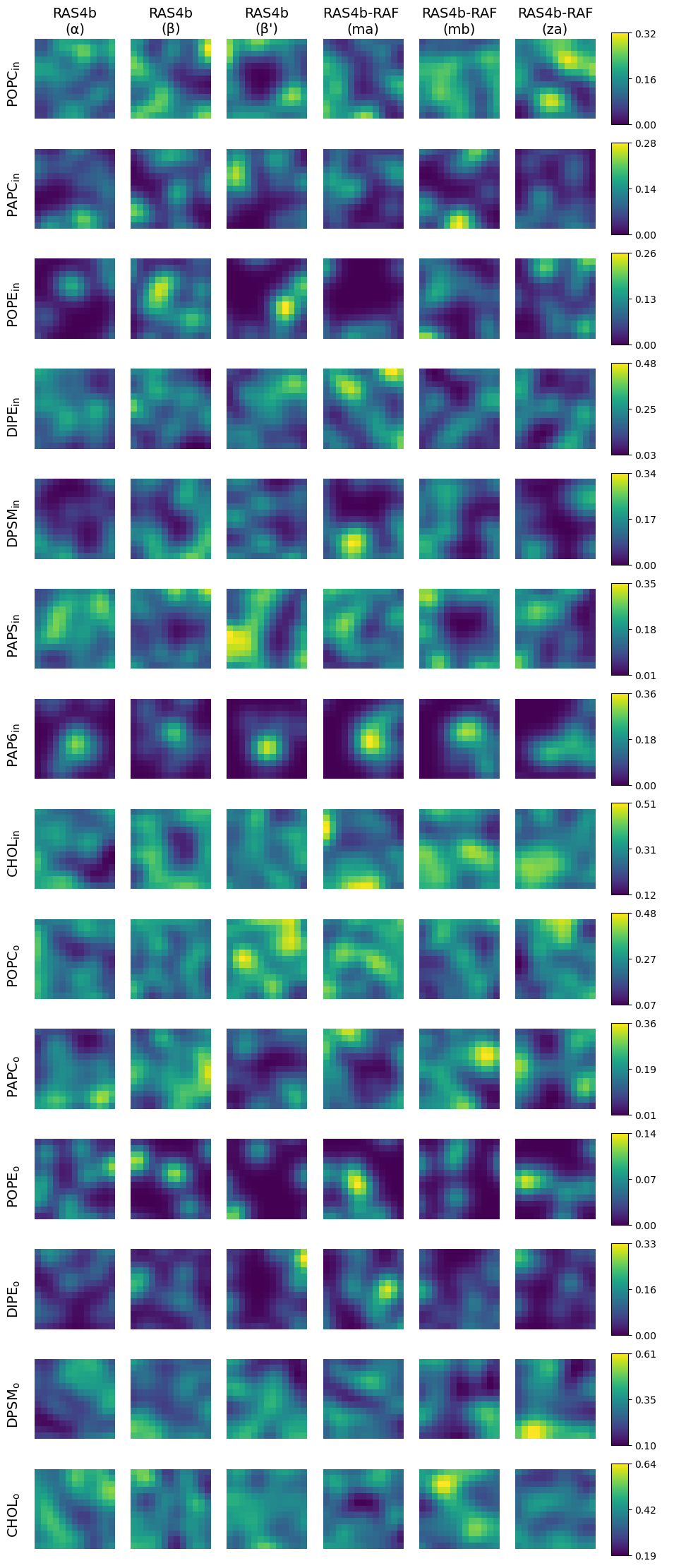}
\vspace{-1.75em}
\caption{Visualization of representative (geometric median of clusters) lipid fingerprints for each of the six protein orientational states highlights the differences between the responses of these lipids. The rows correspond to different types of lipids (and cholesterol), and the columns represent different protein states. The lipid membrane is composed of two layers of lipid molecules (inner and outer monolayers). Lipids PAP6 and PAPS are present only in the inner monolayer of the lipid membrane.
}
\label{fig:patches}
\end{figure}

\autoref{fig:patches} shows data (snapshots) that are representative of clusters for each protein orientational state. Our model learns the spatial features of lipid densities corresponding to each protein state using the data, allowing us to study canonical protein-lipid fingerprints. Differences among such canonical lipid distributions (images) between the six RAS and RAS-RAF orientational states demonstrate that distinct orientations are associated with unique lipid fingerprints. For example, stronger interactions between PAP6 lipids and proteins are observed for protein orientational states \texttt{ma} and \texttt{mb}. Furthermore, it is interesting to note that the protein states having the strongest signals from PAP6 lipids correspond to those with highest accuracies (\texttt{ma} = 90.7\% and \texttt{mb} =91.8\%). These findings suggest that PAP6 lipids have potential to stabilize orientations of proteins in a specific way with respect to the lipid membrane. The geometric median of cluster analysis demonstrates that distinct lipid fingerprints exist for the RAS orientational states, even though some lipid species can be more critical for certain protein orientational state transitions (e.g. change in number of PAP6 lipids around the RAS G-domain). On the other hand, certain lipid species are more common among protein orientational states, and they interact similarly with the RAS and RAS-RAF proteins.

\section{Discussion and Conclusion}

In this work, protein orientational states of RAS and RAS-RAF protein complexes were successfully predicted based on lipid fingerprints from MD simulations, using a ResNet-inspired deep neural network. Our model can predict six protein states based only on their lipid fingerprints with an accuracy of 80\%. A striking relation between the RAS and RAS-RAF orientational states and lipid densities around the protein domains was demonstrated. Since some protein states have low prediction accuracy due to limited training data, there is a potential to increase overall accuracy of the model even further by generating more training data and/or data augmentaion for those states. 

The findings from this work demonstrate the strength of lipid-protein coupling, can be used to improve our understanding of effects of membrane composition on the behavior of RAS proteins and RAS-RAF protein complexes and enable the design of novel therapeutics targeting these mechanisms associated with cancer development. For instance, identification of relation between distinct lipid fingerprints and protein states can enable designing small drug molecules that can regulate protein-lipid interactions so that RAS, RAS-RAF can be stabilized in specific states to prevent undesired functions.


\paragraph{Acknowledgements.}
This work has been supported by the Joint Design of Advanced Computing Solutions for Cancer (JDACS4C) program established by the US Department of Energy (DOE) and the National Cancer Institute (NCI) of the National Institutes of Health (NIH). This work was performed under the auspices of the US DOE by Lawrence Livermore National Laboratory under contract DE-AC52-07NA27344. Release number: \hbadd{LLNL-ABS-834977}.


\clearpage
\bibliography{refs}
\bibliographystyle{icml2022}



\end{document}